# Self-oscillations in field emission nanowire mechanical resonators: a nanometric DC-AC conversion

*Anthony Ayari,[†]\* Pascal Vincent,[†]\* Sorin Perisanu,[†] May Choueib,[†] Vincent Gouttenoire,[†] Mikhael Bechelany[‡], David Cornu,[‡] Stephen T. Purcell[†]*

Université de Lyon, F-69000, France; Univ. Lyon 1, Laboratoire PMCN; CNRS, UMR 5586; F69622 Villeurbanne Cedex, Laboratoire Multimatériaux et Interfaces, Université Lyon 1, CNRS, UMR 5615, Domaine Scientifique de la Doua, F-69622, Villeurbanne Cedex, France

anthony.ayari@lpmcn.univ-lyon1.fr or pascal.vincent@lpmcn.univ-lyon1.fr

We report the observation of self-oscillations in a bottom-up nanoelectromechanical system (NEMS) during field emission driven by a constant applied voltage. An electromechanical model is explored that explains the phenomenon and that can be directly used to develop integrated devices. In this first study we have already achieved ~50% DC/AC (direct to alternative current) conversion. Electrical self-oscillations in NEMS open up a new path for the development of high speed, autonomous nanoresonators, and signal generators and show that field emission (FE) is a powerful tool for building new nano-components.

Within the nanodevice world, nanoelectromechanical systems (NEMS) based on resonant components are having a revolutionary impact on basic research in nanomechanics[1,2] and in technological applications such as wireless communications[3]. Currently the resonators are passive elements in that they need external AC driving signals to enter into resonance, i.e. they are not used to generate independent signals. As well the driving signal is so large that it dominates the output signal from the resonator and makes readout very challenging especially as the device size is reduced to the nanoscale. Until now, driving a nanocantilever by a DC signal or their use as an independent AC source in NEMS has not been achieved and this severely limits integration requiring careful architecture to bring an external AC signal close to the resonator. These limitations are becoming a real bottleneck as researchers orient towards frequencies reaching the GHz.

Clearly it would be an enormous advantage to generate these signals directly in the device, as for example in a Gunn diode[4] which is an illustration of a generic phenomenon called self-oscillation. Self-oscillation, as opposed to the ubiquitous driven harmonic oscillator, is the creation of periodic variations by a constant driving force. This counter-intuitive and intriguing phenomenon appears in a wide range of fields such as physiology, biology, hydrodynamics and electronics. Respective examples are the heart beat,[5] hair bundling in the inner ear,[6] Rayleigh-Benard thermal convection[7] and negative differential resistance electrical devices.[4] In this report, we demonstrate the realization of a bottom-up nanoresonator in the form of a SiC nanowire that generates a tunable AC signal by self-oscillation during field emission from the nanowire apex. Furthermore we go on to identify the system parameters that control the oscillations and fully develop a model that simulates well the experiments.

The self-oscillating resonators consisted of cubic mono-crystalline SiC nanowires attached with a bottom-up method to tungsten tip supports (see ref. 8 for details about the mounting method). These nanowires are promising candidates for NEMS nanocantilevers due to their low density of defects and large modulus to density ratio. They were produced by a vapour-solid process allowing fabrication of large quantities of SiC-based nanowires with diameters $\phi \sim 20\text{-}400$ nm.[9,10] Nanotubes and nanowires are especially useful for FE due to their high aspect ratio that creates a large field at the tip for a relatively low applied voltage. Recently the potential of FE nanosources for high frequency applications has been explored for carbon nanotube microwave diodes[11] of interest in satellite communications, nano-electro mechanical single-electron transistor,[12] and ultra fast pulsed electron sources.[13] We chose a simple single clamped configuration for its versatility, ideal in the exploration of new phenomena in NEMS. This approach allowed the first observation of driven electromechanical resonances of carbon nanotubes in a transmission electron microscope[14] and the first tunable single clamped nanoresonator[15] which were both later exploited in a double-clamped hybrid (top-down/bottom-up) device.[16] FE self oscillations were studied in two experimental setups: (1) an ultra high vacuum (UHV) chamber equipped with FE microscopy; (2) a scanning electron microscope (SEM) equipped with a nanometric manipulator. Measurements were performed on 6 different nanowires. One sample labelled NW1 is shown in the inset of Figure. 1a.

The FE UHV system allows a very stable emission current from the nanowire. Figure 1a is a schema indicating the tip, DC and AC voltage sources ($V_{DC}$, $V_{AC}$) and the phosphor screen on which a FE pattern is formed by the field emitted electrons that accelerate away from the nanowire apex. The pattern consists of several bright spots coming from protrusions or nanometric-scale roughness on the apex which create enhanced local electric fields. Four nanowires (NW1-4) were studied in this setup. Application of the additional AC voltage allowed us to excite the mechanical resonances in which case the entire patterns enlarged in one direction and the emitted current varies.[15] As well the large static electric field creates an axial tension $T \propto V_{DC}^2$ that tunes the resonance frequency which we will see below is important for the self-oscillations. The key moment in the experiment occurred when on increasing $V_{DC}$ with the AC source disconnected, the nanowire image suddenly stretched in one direction (Figure 1d) exactly as observed under the AC excitation at the resonance frequency (Figure 1a). The oscillation was accompanied by a decrease in the average emission current $<I_{FN}>$ as shown in Figure 1b. (The picoammeter has a 1 Hz response time and thus only measures time-averaged signals). Because only a DC voltage was applied we have achieved self-oscillation during FE. The current jumps and hysteresis occurred reproducibly during voltage scans although the onset might change slightly. It is always in the same sense: the current decreased (increased) when the $V_{DC}$ was increased (decreased).

This FE self-oscillation is the main experimental result and the rest of the article is dedicated to its comprehension.

The SEM with nanomanipulator allows to modify *in situ* the geometry of the system and to directly visualise the position and the motion of the nanowire.[17,18] The nanowire was manoeuvred in the vicinity of a metallic sphere that acted as an anode (Figure 2a). The anode or the tip was connected to either DC or AC sources as for the FE system. In AC mode the images widened when the frequency was at the mechanical resonances permitting a measure of the stiffness and quality factor. Application of $V_{DC}$ could bend the nanowire if it was placed asymmetrically with respect to the centre of the anode as shown in Figure 2a. Samples NW5 and NW6 have been studied in this setup.

FE was induced from the nanowire apex to the anode for high enough negative $V_{DC}$ applied to the tip. A DC $I_{FN}$ was then detected on the picoammeter which generally correlated with an increase in the image intensity, even up to saturation (Figure 2b-c). The SEM uses a secondary electron detector (SED) for imaging and the increased intensity is due to extra secondary electrons created at the sphere by the nanowire FE current. $V_{DC}$ used here was ~100V for a $\phi \approx 100$ nm but could be decreased to tens of volts by approaching the nanowire to the anode or by using a lower diameter nanowire.[18] We used a large diameter sample in this experiment to ease imaging, limit the influence of the SEM electron beam and lower the resonant frequency, as our nanowires have a fairly constant aspect ratio. On increasing $V_{DC}$ (and $I_{FN}$) with the AC generator disconnected, the nanowire image suddenly widened (Figure 2c) showing that it had again jumped into self-oscillations as in the UHV system. To check that this vibration was not induced by the SEM beam, we turned off the scanning and moved the beam away from the nanowire. $V_{DC}$ was then again swept and the SED signal recorded on an oscilloscope. For the same voltage and current range as before an AC signal was measured (Figure 2d), confirming the oscillating motion of the nanowire while emitting electrons.

Constructing a viable model depends on identifying and estimating the different elements of the electromechanical device that may come into play. These include the Young's modulus E and the quality factor Q of the nanocantilever, the applied DC voltage at the tip end, the charge on the nanowire apex, the electrical resistivity of the nanowire, the emission current $I_{FN}$ that also transits into the nanowire, and the nanowire/anode capacitance. The flexibility of the SEM system allows us to measure most of these parameters directly. It is straightforward to measure E, Q (~4,000), and $I_{FN}$ (pA-100 nA). The nanowire resistances were measured to be $10^9$-$10^{10}$ ohms which is typical for this rather large gap semiconductor. Two methods were used: (1) two point IV measurements directly in the SEM by touching the end of nanowire to the sphere; (2) analysis of the electron energy spectra of the FE electrons in the UHV system as previously carried out for carbon nanotubes.[19] Further useful information is gained by comparing the behaviour for the two voltage polarities. With positive polarity on the nanowire there is no emission for our voltage range because of the large sphere radius and hence low field enhancement. No nanowire vibrations then occurred showing that the FE current was essential for self-oscillations. A more subtle aspect which turns out to be critical for the phenomenon is unearthed when one compares the resonance frequencies for opposite polarities determined by AC excitation as a function of $|V_{DC}|$. The frequency for zero $V_{DC}$ gave E= 500 GPa which is typical for this high quality nanowire. Figure 2e shows a parabolic behaviour of the resonant frequency due to the electrostatic tension as observed previously.[15,16] For negative $V_{DC}$ below the emission threshold, the resonant frequencies coincided with those for positive polarity. However above the emission threshold we turned off the AC signal as well as the electron beam scanning and measured the self-oscillation frequency with the SED signal. We then observed a clear departure from the positive bias parabola. This is because now the voltage at the apex which is responsible for the frequency tuning is no longer the same as $V_{DC}$ due to the voltage drop along these high resistance nanowires. A final important point for the model, and also for DC/AC conversion, is the variation in $I_{FN}$ during oscillation. From the SED signal in Figure 2d, we get an AC amplitude of 3V while the average DC voltage is 6 V higher than the one without field emission. So, we can roughly estimate that the AC component can be as high as fifty percent of the DC current, which is excellent for a non-optimised geometry.

Previously, torsional self-excitation by mode coupling[20] and a laser driven self resonant NEMS where the driving mechanism is cyclic heat dilatation[21], have been realized with a light-reflecting, flat cantilever in an interferometer arrangement. The later approach is inconvenient for an integrated device and high frequency application compared to a purely electrically driven self resonant NEMS. Thermal effects, as well as mode coupling, can be ruled out for our samples. With the measured voltage drop across the nanowire, the current and a thermal conductivity of 1 W m$^{-1}$ K$^{-1}$ (rather pessimistic for a SiC single crystal), we estimate an increase of temperature below 2 degrees. Moreover, at this temperature the amplitude of the thermal oscillations such as those observed in ref. 22 is lower then 5 nm, much below the amplitude observed on Figure 1d and Figure 2c. We show next that a model in which only electromechanical effects are involved explains our experiments.

The motion of the nanowire apex can be described as an oscillator with a variable resonance frequency due to the electrostatic tension $T = f(x)V^2$, submitted to a bending force $F_b = g(x)V^2$ and with a phenomenological damping factor $\omega_0/Q$. We make the approximation that $T$ and $F_b$ are concentrated at the apex:

$$\ddot{x} + (\omega_0^2 + bT)x = \frac{F_b}{m_{eff}} - \frac{\omega_0}{Q}\dot{x} \quad (1)$$

The x axes is represented in Figure 2a, $\omega_0$ is the fundamental frequency at $V_{DC} = 0$, $m_{eff}$ the effective mass, f(x) and g(x) two functions that depend on the capacitive environment, $V$ the voltage at the nanowire apex and b is a constant related to the mass and length of the nanowire.

Using Kirchhoff's law, the electrical circuit of this NEMS is governed by the following equation:

$$I_{FN} = D(\beta V)^2 \exp(-\frac{d}{\beta V}) = \frac{V_{DC} - V}{R_{NW}} - \frac{d}{dt}(CV) \quad (2)$$

$D$ and $d$ are constants, β(x) the field enhancement factor (related to the nanowire radius and length and to the distance to the anode), $R_{NW}$ the resistance of the nanowire and $C$ the capacitance between the grounded anode and the apex. The last term accounts for the cyclic variation of charge.

It is beyond the scope of this work to determine β(x), bf(x), g(x)/m$_{eff}$ and C(x), by finite element simulations. Instead, we performed numerical simulation on simplified equations that can be solved analytically and compared with simulation, with no loss of generality and less unknown parameters. The analytical and semi analytical resolution is very useful to prove the absence of numerical instability. The simplified equations are: $\ddot{X}+\frac{\omega_0}{Q}\dot{X}+\omega_0^2(X-x_0)+p^2V^2X=0$ with X= $x_0$-x and

$-I_0/(1+aX^2)+\frac{V_{DC}-V}{R_{NW}}=C\dot{V}$ with $I_0=D\beta_0^2V^2\exp(-\frac{d}{\beta_0V})(1+ax_e^2)$ and $x_e$ the equilibrium position for a given $V_{DC}$, $x_e=\frac{x_0\omega_0^2}{p^2V^2+\omega_0^2}$. The approximate form of the emission current takes into account the boundary condition and the fact that the nanowire is not perfectly aligned with the anode, i.e. that it bends upon the application of $V_{DC}$. The parameters used in the following are for NW6. This gives $x_0$= 20 µm (obtained from Figure 2a), $D(\beta_0)^2$ = 2.2 $10^{-9}$ AV$^{-2}$, $d/\beta_0$ =2276.5 V, $R_{NW}$=5.10$^9$ ohm, $p^2$= 2178576 rad$^2$s$^{-2}$V$^{-2}$ (obtained from Figure 2e), $\omega_0$=2*π*42150 rad/s, Q=4000, C=10$^{-17}$ F, a=10$^{11}$m$^{-2}$. C and a are the only unknown parameters. The value of C used here is close to the self capacitance of a sphere of the same diameter as the nanowire. a is chosen such that $aX^2$ is comparable to one for small amplitude oscillations.

For the analytical study we injected X=$x_e$+r(t)cos(ωt) (where r(t) is the slowly varying oscillation amplitude), V=$U_e$+$U_0$cos(ωt+φ)+$U_1$cos(2ωt+ψ) and I= $I_0$/(1+aX$^2$) ≈ $I_0$[A($x_e$,r)+B($x_e$,r)cos(ωt)+J($x_e$,r) cos(2ωt)] (where $A=\frac{\omega}{2\pi}\int_0^{2\pi/\omega}\frac{dt}{1+a(x_e+r\cos\omega t)^2}$, $B=\frac{\omega}{\pi}\int_0^{2\pi/\omega}\frac{\cos\omega t}{1+aX^2}dt$ and $J=\frac{\omega}{\pi}\int_0^{2\pi/\omega}\frac{\cos 2\omega t}{1+aX^2}dt$ are Fourier coefficients of I for constant r, ω$^2$=ω$_0^2$+p$^2$V$^2$ and $U_e$, $U_0$ and $U_1$ are some constants) into our equations and we checked the stability by looking at the sign of the sin(ωt) component of the mechanical equation. After some algebra, we get :

$$\frac{dr}{dt}\approx\frac{1}{2}\left\{-\frac{\omega_0}{Q}r-2p^2R_{NW}^2CI_0[x_eB(x_e,r)+rJ(x_e,r)](V_{DC}-I_0R_{NW}A(x_e,r))\right\} \quad (3)$$

This equation gives the equilibrium positions and their stability.

Numerical simulations performed with eqs (1) and (2) and analytical and semi analytical methods confirmed that they give rise to either immobile or self-oscillating solutions depending on the applied voltage (Figure 3). An equilibrium solution exists if eq (3) equals zero and is stable if d(dr/dt)/dr <0.

The analytical approaches showed that the variation of C by several orders of magnitude has little impact on the voltage where self-oscillations appear and the most important ingredients are the x dependence of $\beta$ (and of $I_{FN}$) and the voltage dependence of the resonant frequency, that allows the energy stored in the capacitance to be re-injected into the mechanical oscillator to cancel the energy loss. The existence of oscillations is also facilitated by the exponential dependence in $I_{FN}$ and low damping. Figure 3b shows reasonable qualitative agreement between theory and experiment concerning the needed DC voltage as well as the AC current amplitude (inset). The immobile solution coexists with a self-oscillating one over a certain range of voltage and becomes unstable for high enough voltage in agreement with the hysteresis observed in Figure 1b.

In conclusion we have shown how to create a self oscillating NEMS device. The parameters that come into play are all elements that can be included in integrated devices, as radio frequency micromachined devices using FE have already been fabricated.[23] We are currently pursuing top-down and bottom-up methods with the goal of building a functioning device on these principles. We anticipate that this self-oscillating behaviour will solve one of the major issues of NEMS which is cross talk. The crucial point here is that a tunable self-oscillating NEMS doesn't need an external AC source. This is an important step towards making NEMS active rather than passive devices.

ACKNOWLEDGMENT (Word Style "TD_Acknowledgments"). This work was carried out within the framework of the Group Nanowires-Nanotubes Lyonnais. The authors acknowledge the support of the "Centre Technologique des Microstructures de l'Université Lyon 1". M. C. thanks the Lebanese CNRS for financial support.

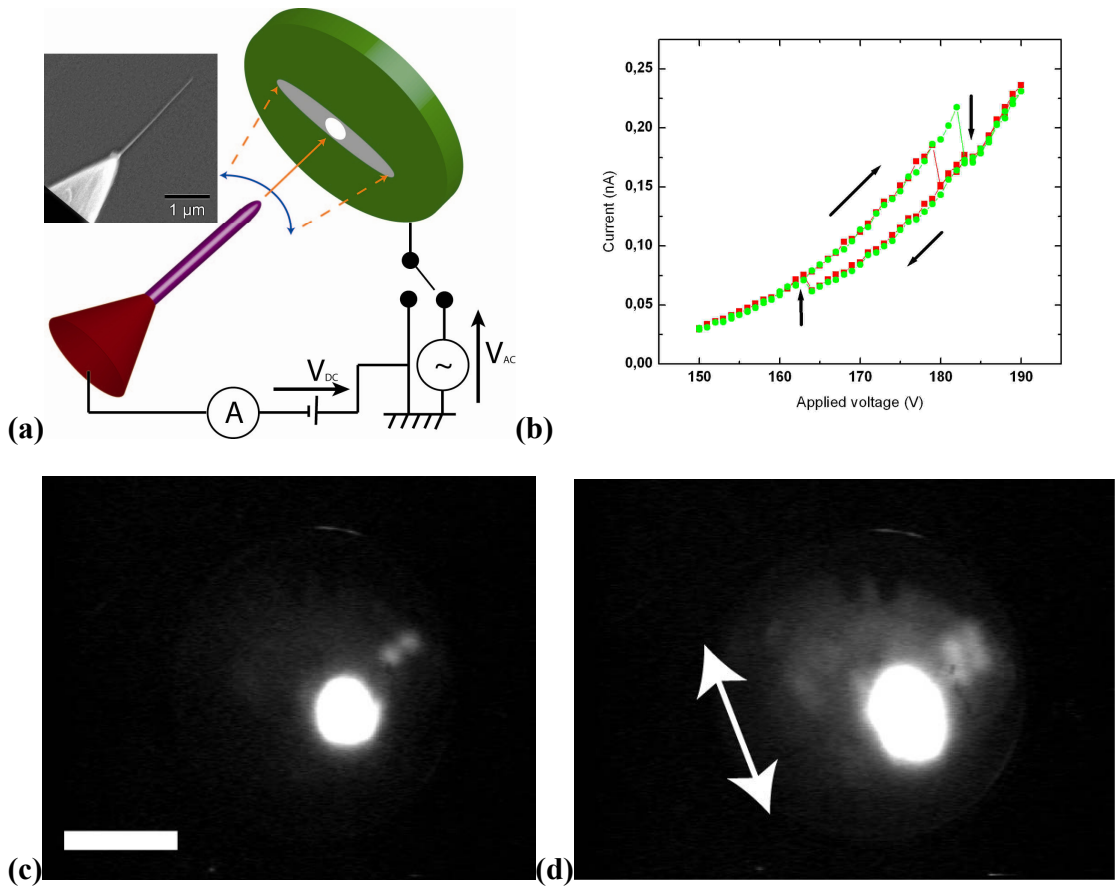

**Figure 1.** (a) Schematic diagram of the UHV system. Inset: SEM image of nanowire NW1 (ϕ = 30 nm, length (L) 2 μm, resonant frequency in the tens of MHz range) mounted on a tungsten tip. (b) Emission current as a function of applied voltage for two successive voltage sweeps for NW3 (ϕ = 200 nm, L=50 μm). The arrows show the direction of sweeping and current jumps. Field emission patterns of the nanowire NW4 (ϕ = 30 nm, L= 20 μm) immobile (c) and in self-oscillation (d). The pattern widens in the direction of motion of the nanowire indicated by the white arrow. As the phosphor screen is 3 cm away from the nanowire and the size of the stretched pattern is about 5 mm (scale bar 1 cm), the amplitude of the oscillation can be estimated to be a few hundred nanometers.

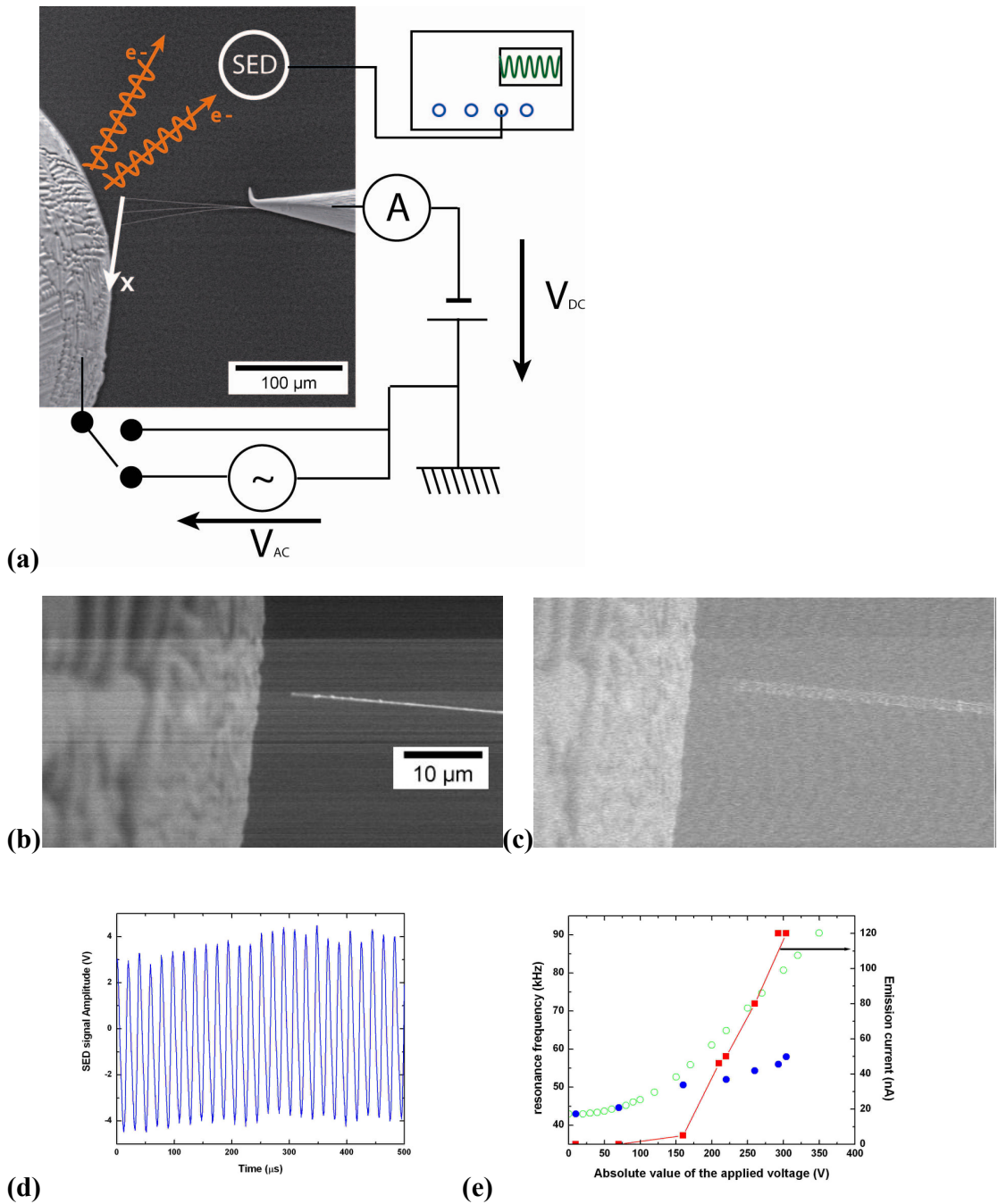

**Figure 2.** (a) Schematic diagram of the SEM system with superimposed, SEM images of NW6 ($\phi$ = 400 nm, L= 125 μm) at different static bending, secondary electron emission and electrical connections to the oscilloscope, AC or DC generators. (b) SEM image of NW5 ($\phi$ = 250 nm, L = 270 μm) at rest. (c) The same nanowire in self-oscillation with an amplitude about 1 μm. (d) Signal from the SED for NW6 for $V_{DC}$= 220 V and $I_{FN}$ = 50 nA. The Fourier transform gives a frequency of 54 kHz. The oscilloscope

was in AC coupling in order to remove the DC component. (e) Frequency tuning on NW6 ((o) positive bias, (•) negative bias and (■) emission current).

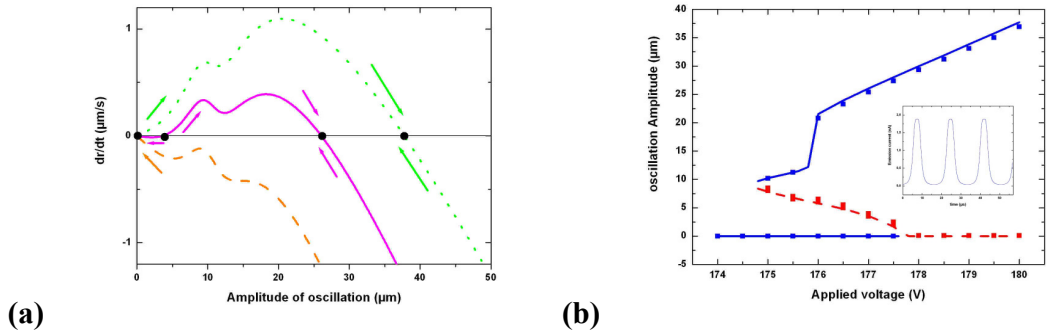

(a) (b)

**Figure 3.** (a) Analytical calculation of the derivative of the amplitude dr/dt over time versus amplitude of oscillation for different applied voltage. The dashed (respectively solid, dotted) line is for $V_{DC}$ = 174 V (respectively 177 V, 180 V). Equilibrium points are shown by filled circles. The arrows indicate the stability of each point. (b) Stability diagram for different applied voltages. The solid lines are the stable branches and the dashed line is the unstable branch obtained from analytical calculations. The squares are obtained from numerical simulations and are in excellent agreement with analytical calculations. Inset: Emission current as a function of time at equilibrium for $V_{DC}$ = 176 V.

SYNOPSIS TOC (Word Style "SN_Synopsis_TOC").

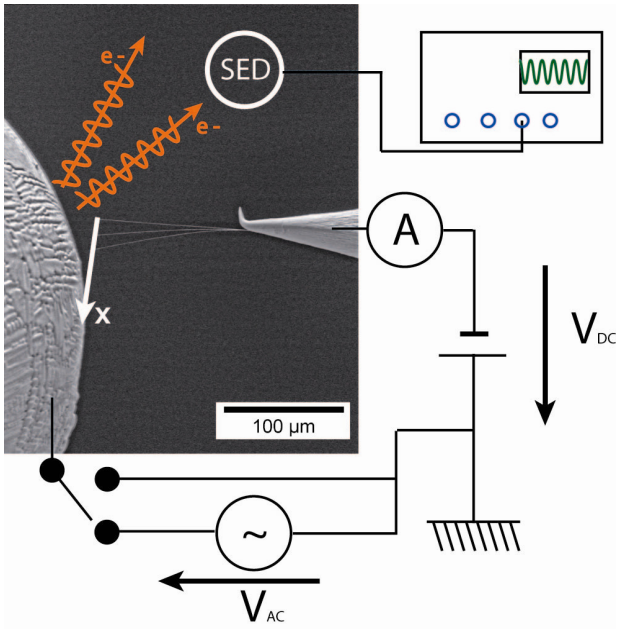